\documentstyle[prb,aps,amsfonts,amssymb,epsf]{revtex}
\begin{document}
\draft 
\twocolumn[\hsize\textwidth\columnwidth\hsize\csname @twocolumnfalse\endcsname

\title{Systematics of $c$-axis Phonons in the Thallium and Bismuth Based Cuprate 
Superconductors}

\author{A. A. Tsvetkov$^{1,2}$\cite{byline}, D. Duli\'c\cite{byline}, 
        D. van der Marel, A. Damascelli}
\address{$^1$ Laboratory of Solid State Physics, Materials Science Center, 
        Groningen University, Nijenborgh 4, 9747 AG Groningen, The~Netherlands}
\author{G. A. Kaljushnaia, J. I. Gorina, N.~N.~Senturina}
\address{$^2$ P. N. Lebedev Physical Institute, Russian Academy of Sciences, 
        Leninsky Prospect 53, 117924 Moscow, Russia}
\author{N. N. Kolesnikov}
\address{Institute of Solid State Physics, Russian Academy of Sciences,
      Chernogolovka 142432, Russia}
\author{Z. F. Ren, J. H. Wang}
\address{Department of Chemistry, SUNY at Buffalo, 
      Buffalo NY 14260-3000, USA}
\author{A. A. Menovsky}
\address{Van der Waals-Zeeman Laboratory, University of Amsterdam, The 
Netherlands}     
\author{T. T. M. Palstra}
\address{Inorganic Solid State Chemistry Laboratory, Materials Science Center, 
        Groningen University, Nijenborgh 4, 9747 AG Groningen, The Netherlands}     
\date{\today}
\maketitle
\begin{abstract}
We present grazing incidence reflectivity measurements in the far infrared 
region at temperatures above and below T$_{\text{c}}$ 
for a series of thallium 
(Tl$_{2}$Ba$_{2}$CuO$_{6}$, 
Tl$_{2}$Ba$_{2}$CaCu$_{2}$O$_{8}$) 
and bismuth 
(Bi$_{2}$Sr$_{2}$CuO$_{6}$,
Bi$_{2}$Sr$_{2}$CaCu$_{2}$O$_{8}$, and
Bi$_{2-\text{x}}$Pb$_{\text{x}}$Sr$_{2}$CaCu$_{2}$O$_{8}$) 
based cuprate superconductors. From the spectra, which are dominated by
the $c$-axis phonons, longitudinal frequencies (LO) are directly obtained. 
The reflectivity curves are well 
fitted by a series of Lorentz oscillators. In this way the transverse (TO)  
phonon frequencies were accurately determined.
On the basis of the comparative study of the Bi and Tl based cuprates with 
different number of CuO$_2$ layers per unit cell,
we suggest modifications of the assignment of the main oxygen modes.
We compare the LO frequencies in Bi$_{2}$Sr$_{2}$CaCu$_{2}$O$_{8}$ and 
Tl$_{2}$Ba$_{2}$Ca$_2$Cu$_{3}$O$_{10}$
obtained from intrinsic Josephson junction characteristics 
with our measurements, and explain the discrepancy in LO frequencies 
obtained by the two different methods.
\end{abstract}

\vskip2pc]
\narrowtext

                             %%%%%%%%%%%%%%%%%%
                             %  Introduction  %
                             %%%%%%%%%%%%%%%%%%

\section{Introduction}
%\narrowtext

The phonon spectra of high temperature superconductors were widely studied 
due to the possible important role in the mechanism of superconductivity. 
Besides that, the phonons being on the same energy scale as the superconducting gap
strongly affected many superconductivity related quantities, such as electronic 
response and the superconducting energy gap, pseudo-gap, Josephson plasma 
oscillations, and tunneling spectra.
The first optical measurements were performed on polycrystalline 
samples.\cite{Litvinchuk} The assignment of the phonon modes and 
the lattice dynamical calculations used for the assignment were strongly 
influenced by the mixed in-plane and out-of-plane response in these earlier data. 
As large single crystals became available, the infrared (IR) properties and, 
in particular, the $c$-axis phonons of the yttrium and lanthanum based compounds 
were thoroughly investigated, providing phonon eigenvector 
patterns\cite{Schuetzmann95,Homes,JKim,Henn} different from the early
assignment. The $c$-axis response of 
the bismuth and thallium classes of superconductors is known to 
a much lesser extent. The measurements done on these materials 
probed primarily the $ab$-properties,\cite{Zibold,Terasaki,Puchkov,Artem1} 
or polycrystalline samples were 
used.\cite{Popovic,Foster,Zetterer,Kamaras,Mihajlovic}
The unambiguous information on the IR $c$-axis phonon spectra were obtained 
on Bi$_{2}$Sr$_{2}$CaCu$_{2}$O$_{8}$ (Bi2212) single crystals\cite{Zibold1,Tajima}
and on 
Tl$_{2}$Ba$_{2}$Ca$_2$Cu$_{3}$O$_{10}$ (Tl2223) film.\cite{Kim}
Extensive studies, along with the efforts to assign
the main oxygen modes in these compounds were also performed using Raman 
scattering.\cite{Gasparov,Kall2,Liu,Nemetschek,Kall1,Krantz,Kakihana,Kang,Kendziora,Martin,Osada,Chen}

In this paper we present a systematic study of the $c$-axis IR phonons in Bi and
Tl cuprates by means of grazing angle reflectivity measurements. We discuss 
the technique used and show its sensitivity to the $c$-axis properties 
in anisotropic materials. 
The comparison with the available data on Bi2212 by 
Tajima {\it et al.}\cite{Tajima} proves that both transverse (TO) and 
longitudinal (LO) $c$-axis frequencies can be determined correctly by this method. 
We report on the first measurements of the $c$-axis vibrational spectra 
in single layer Bi and Tl compound, done on single crystals. 
By comparing bismuth and thallium compounds with the different number 
of CuO$_2$ layers per unit cell (single and double layer systems), 
we propose a consistent assignment of the main oxygen modes, 
different from those done by lattice dynamical calculations.\cite{Prade,Kulkarni}
In the bismuth cuprate, using lead-doped samples, we
identify the superstructure induced modes. 
The particular sensitivity of our technique to the longitudinal modes 
enables a direct comparison of our data to the fine
structure observed in the intrinsic Josephson junction 
characteristics,\cite{Schlenga2,Yurgens}
which was ascribed to the LO phonon excitations.\cite{Helm}

                       %%%%%%%%%%%%%%%%%%%%%%%%%%%%%%%
                       %       Experimental          %
                       %%%%%%%%%%%%%%%%%%%%%%%%%%%%%%%

\section{Experimental}

The reflectivity measurements were performed at a grazing angle of 80$^\circ$
relative to the surface normal, with p-polarized light (electric vector in 
the plane of incidence), in the far-infrared (FIR) region (30-700 cm$^{-1}$).
Reference spectra were obtained by evaporating a gold film {\it in situ}
on the sample surface, and absolute reflectivities were determined by 
reflectivity ratio. 
A wire grid polarizer was used to select the proper polarization. 
The measurements were made using  a Fourier transform 
spectrometer (Bruker IFS 113v). A series of samples was measured: 

Bi$_{2}$Sr$_{2}$CuO$_{6}$ (Bi2201),
Bi$_{2}$Sr$_{2}$CaCu$_{2}$O$_{8}$ (Bi2212) and lead doped 
Bi$_{2-\text{x}}$Pb$_{\text{x}}$Sr$_{2}$CaCu$_{2}$O$_{8}$ (Bi2212+Pb) 
single crystals were grown at the Lebedev Physical Institute by 
the method of re-crystallization from the precursor dissolved 
in melted KCl.\cite{Gorina1,Gorina2,Martovitsky}
As-grown samples were obtained as free-standing thin crystalline platelets and 
did not require any further treatment. 
The transition temperatures (and the transition widths) were
T$_{\text{c}}$= 6-8 K ($\Delta$T$_{\text{c}}=2$ K), 
T$_{\text{c}}=80$ K ($\Delta$T$_{\text{c}}=2$ K), and 
T$_{\text{c}}=75$ K ($\Delta$T$_{\text{c}}=3$ K), respectively. The lead 
contents in doped
samples was $\text{x}=0.4$, as determined by the x-ray electron probe 
microanalysis. 

Bi$_{2}$Sr$_{2}$CaCu$_{2}$O$_{8}$ single crystals of a larger size 
($3\times4$ mm) were grown at the University of Amsterdam by the travelling 
solvent floating zone method.\cite{Menovsky} 
The transition temperature and the transition width were T$_{\text{c}}=89$ K 
and $\Delta$T$_{\text{c}}=3$ K, respectively.
An optically clean surface was obtained by peeling the samples along 
the ab-plane.

Tl$_{2}$Ba$_{2}$CuO$_{6}$ (Tl2201) single crystals were produced at
the Institute of Solid State Physics. The samples, with T$_{\text{c}}=82$ K 
and $\Delta$T$_{\text{c}}=13$ K, were obtained by the self-flux 
method.\cite{Kolesnikov1,Kolesnikov2}

Tl$_{2}$Ba$_{2}$CuO$_{6}$ films came from SUNY at Buffalo. 
They were made by radio-frequency magnetron sputtering followed 
by post-deposition annealing.\cite{Wang}
The films had T$_{\text{c}}=80$ K. 
More details on the preparation and properties of 
the films can be found in Refs.~\onlinecite{Wang} and \onlinecite{Ren}.

Tl$_{2}$Ba$_{2}$CaCu$_{2}$O$_{8}$ (Tl2212) films were provided by 
Superconductor Technologies Inc. and were produced using their standard production 
methods \cite{MMEddy96p} and had $T_c = 98 {\rm{~K}}$. More details on the
electrical properties of these films can be found in \onlinecite{Willemsen}.

The samples were mounted in a conventional liquid He flow optical cryostat,
which allowed any temperature, between room temperature and 8 K, to be reached. 

                        %%%%%%%%%%%%%%%%%%%%%%%%%
                        %       TECHNIQUE               %
                        %%%%%%%%%%%%%%%%%%%%%%%%%

\section{Technique}
The geometry of the grazing angle technique is shown in the inset of 
Fig.~\ref{Reflectivity}. 
From the figure it is easy to see that s-polarized 
light probes the $ab$-plane  properties of a crystal, whereas p-polarization 
gives much information about the $c$-axis response. 
This is a convenient method for extracting both the $ab$-plane 
($\varepsilon_{ab}=
\varepsilon_{ab}^{\prime}+i\/\varepsilon_{ab}^{\prime\prime}=
n_{ab}^2$), 
and the $c$-axis 
($\varepsilon_{c}=\varepsilon_{c}^{\prime}+i\/\varepsilon_{c}^{\prime\prime}$) 
components of the dielectric tensor, in case normal incidence measurements on 
the $ac$-plane are not possible due to the small size of samples 
along the $c$-axis, or to the mica-like nature of the Bi compounds.
All cuprate superconductors are highly anisotropic in the $ac$, or $bc$-planes.
The $ab$-plane anisotropy is much smaller. It does not affect our
analysis and is neglected in the following discussion. 
In this case the reflectivity $R_{p}$ is described by the Fresnel formula for 
a uniaxial crystal:
\begin{equation}
\label{eq:fresnel}
R_p = \left|\frac{\sqrt{\varepsilon_{ab}}\cos\theta - 
\sqrt{1-\frac{\sin^2\theta}{\varepsilon_c}}}
{\sqrt{\varepsilon_{ab}}\cos\theta +
\sqrt{1-\frac{\sin^2\theta}{\varepsilon_c}}}\right|^2 ,
\end{equation}
where $\theta$ is the angle of incidence, in this case equal to 80$^\circ$. 
In order to fit the data, we need to assume a form for $\varepsilon_{ab}$ and
$\varepsilon_{c}$. 
The most common procedure is to use the Drude or two-fluid 
model (in case of a superconductor):
\begin{eqnarray}
\label{eq:difun}
\varepsilon_{ab,c}(\omega)  =  
 \varepsilon_\infty^{ab,c}  - \frac{\omega_{p_{ab,c}}^2}{\omega^2} &&+ 
 \frac{4\pi i \sigma_{ab,c}}{\omega}\nonumber\\
 &&+ \sum_j \frac{S_j\omega_{T_{j}}^2}{\omega_{T_{j}}^2-\omega^2-
i\omega\gamma_j},
\end{eqnarray} 
where the first term represents high energy interband transitions, 
the second and the third terms are superconducting and normal carrier 
responses, respectively, and the last one is a sum of phonon contributions. 
In the normal state, the electronic background largely affects
$\varepsilon_{ab}$, since the $ab$-planes are metallic, whereas 
the $c$-axis properties are insulator-like. 
This means that the most prominent phonon modes are polarized along 
the $c$-axis, with the $ab$-plane phonons being strongly screened by 
the free carriers.
The reflectivity function $R_{p}$ has sharp minima when
$\varepsilon_{c}^{\prime}$ goes to zero, i.e. at the longitudinal frequencies, 
while the general trend depends on $\varepsilon_{ab}$. 
The phonons associated with the $ab$-plane are much broader and almost absent
in the spectra, so we can assume that $\varepsilon_{ab}$ is a smooth function
of frequency.

The advantage of the grazing angle technique is that the $c$-axis longitudinal 
frequencies are determined directly from the measured data, without 
any modelling. 
From the reflectivity we can calculate a pseudo-loss function, $L(\omega)$,
defined as\cite{houston}:  
\begin{equation}
\label{eq:pseudo}
L(\omega)=
\frac{(1-R_p)|n_{ab}|\cos\theta}{2(1+R_p)}\sim
\text{Im}\,e^{i\phi}\sqrt{1-\frac{\sin^2\theta}{\varepsilon_c}},
\end{equation}
where $n_{ab}$ is the in-plane complex refraction index and 
$\phi = \pi/2-\arg{(n_{ab})}$ is a weakly frequency dependent phase shift. 
$L(\omega)$ is approximately the same as the $c$-axis loss-function 
$\text{Im}(-1/\varepsilon_{c})$ (at least the maxima of both functions 
are at the same positions), and  it gives us the frequencies of 
the $c$-axis longitudinal phonons. 
The reason for observing the LO phonon (and not TO)
frequencies in $R_p$ is common for the $c$-axis plasmon\cite{Artem2} and 
phonons, and is the following:
The incident electromagnetic radiation excites a wave inside the material, 
which has an electric field component
${\bf\vec{\tilde{E}}}={\bf\vec{E}}e^{i({\bf\vec{k}}{\bf\vec{r}}-\omega t)}$, 
where each component of ${\bf\vec{E}}$ and ${\bf\vec{k}}$ can be complex. 
If the {\it x}-axis lies at the intersection of the plane of incidence with
the surface, and the {\it z}-axis is normal to the surface, then 
the {\it x}-component of the wave vector, $k_x$, is determined by 
the incident radiation, while the {\it z}-component of the wave vector 
can be shown to depend both on the in-plane and out-of-plane dielectric 
functions, as follows:
\begin{equation}
\label{eq:wavevector}
k_x = \frac{\omega}{c}\sin\theta, \quad\quad\quad
k_z^2 = \frac{\omega^2}{c^2}\varepsilon_{ab}(1-
\frac{\sin^2\theta}{\varepsilon_c})
\, ,
\end{equation}
where $\omega$ is the frequency of incident radiation and $c$ is 
the speed of light.
Inside the medium the wave remains polarized in the {\it xz}-plane, and 
the components of electric field relate to each other as:
\begin{equation}
\label{eq:field}
\frac{E_z}{E_x} = \frac{\sqrt{\varepsilon_{ab}}}{\sqrt{\varepsilon_c - 
\sin^2\theta}}.
\end{equation}
This illustrates the well-known fact that the electromagnetic wave in 
an anisotropic medium does not have a pure transverse character, but 
it has also a longitudinal component, unless it propagates
along the principal axes.

In most high-T$_c$ compounds the {\it c}-axis plasma frequency is far below 
the optical phonon frequencies and the phonon modes are well separated 
in frequency, while the in-plane response is mostly dominated by plasma 
oscillations of carriers in the {\it ab}-plane. 
Thus, the dielectric function $\varepsilon_{ab}(\omega)$ 
at low frequencies can be written as a pure plasma response, and 
$\varepsilon_{c}(\omega)$ near a particular phonon can be simplified to 
a single Lorentz oscillator:
\begin{equation}
\label{eq:plasma}
\varepsilon_{ab}(\omega) = \varepsilon_\infty^{ab} - 
\frac{\omega_{pab}^2}{\omega^2},
\quad\quad
\varepsilon_{c}(\omega) = \varepsilon_{sc} 
+ \frac{S\omega_{T}^2}{\omega_{T}^2-\omega^2},
\end{equation}
where $\omega_{pab}$ is the in-plane plasma frequency, $\omega_{T}$ is 
the TO phonon frequency, $S$ is the oscillator strength,
and $\varepsilon_{sc}$ includes also contribution to $\varepsilon_{\infty}$ 
from other phonons. In both functions, we neglect the damping for simplicity.

The substitution of Eqs.~\ref{eq:plasma} into Eqs.~\ref{eq:wavevector} gives 
a pure imaginary solution for $k_z(\omega)$ everywhere, except for 
a narrow frequency range, where $\omega^2$ varies between 
$\omega_{L}^2=\omega_{T}^2+S\omega_{T}^2/{\varepsilon_{sc}}$
and $\omega_{T}^2+S\omega_{T}^2/(\varepsilon_{sc}-\sin^2\theta)$. 
In this region $k_z(\omega)$ is real. 
This corresponds to the propagation of a travelling electromagnetic wave in 
the medium, which reduces the reflectivity. 
Outside this narrow frequency range the electromagnetic 
wave decays exponentially near the surface, and the total 
reflection takes place. 
The finite scattering rate in the planes gives rise to the Hagen-Rubens behavior 
of the reflectivity, while the phonon damping together with the $c$-axis 
conductivity determine the width of the phonon lines.

                        %%%%%%%%%%%%%%%%%%%%%%%%%
                        %       RESULTS         %
                        %%%%%%%%%%%%%%%%%%%%%%%%%

\section{Results}
\begin{figure}[t]
\epsfxsize=14cm
\centerline{\epsffile{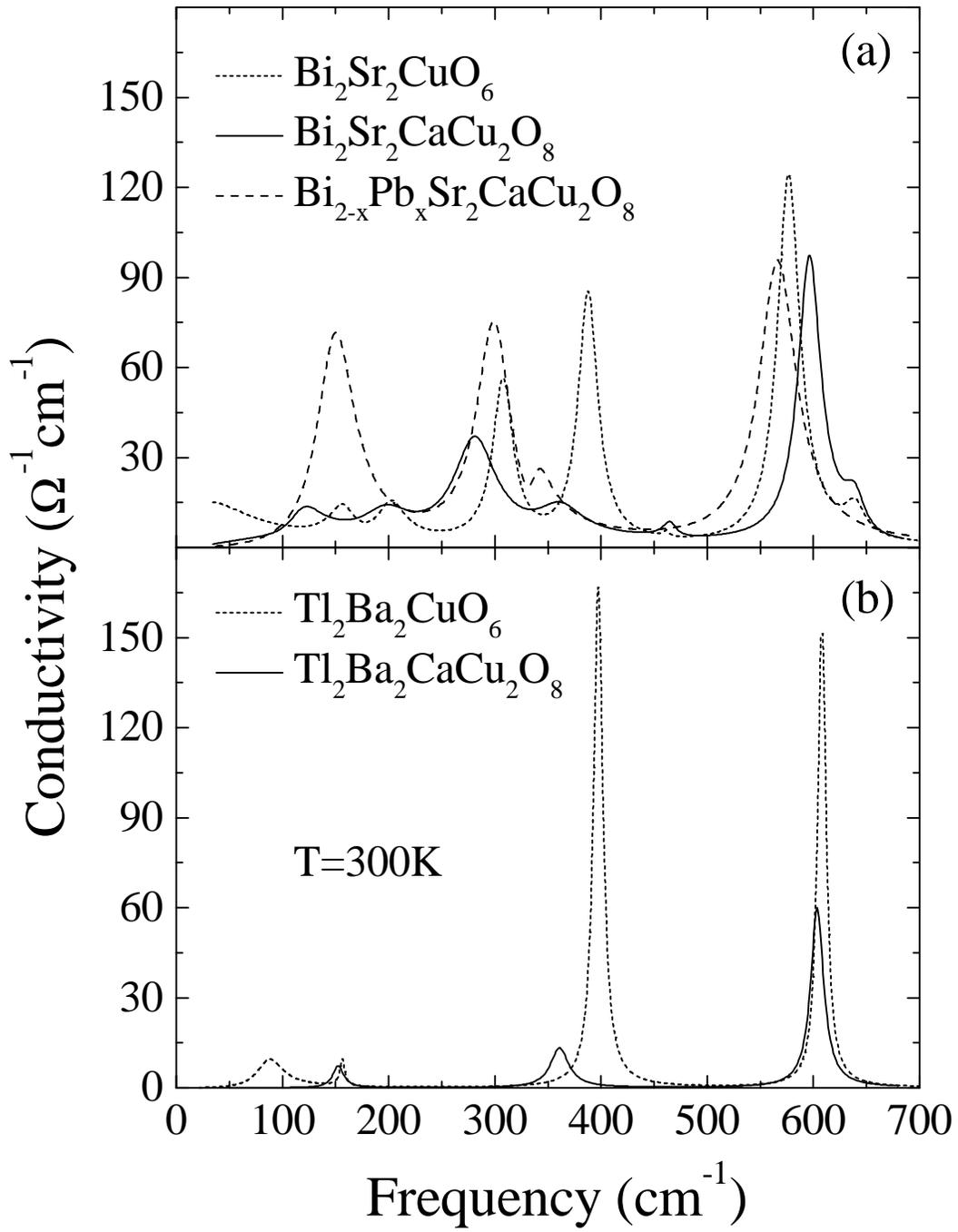}}
\vspace{20mm}
\caption{The $c$-axis conductivity (a) for the Bi and
(b) for Tl compounds, obtained from the fit of 
the reflectivity to the Drude-Lorentz model.}
\label{Conductivity}
\end{figure}

\begin{figure}[t]
\epsfxsize=14cm
\centerline{\epsffile{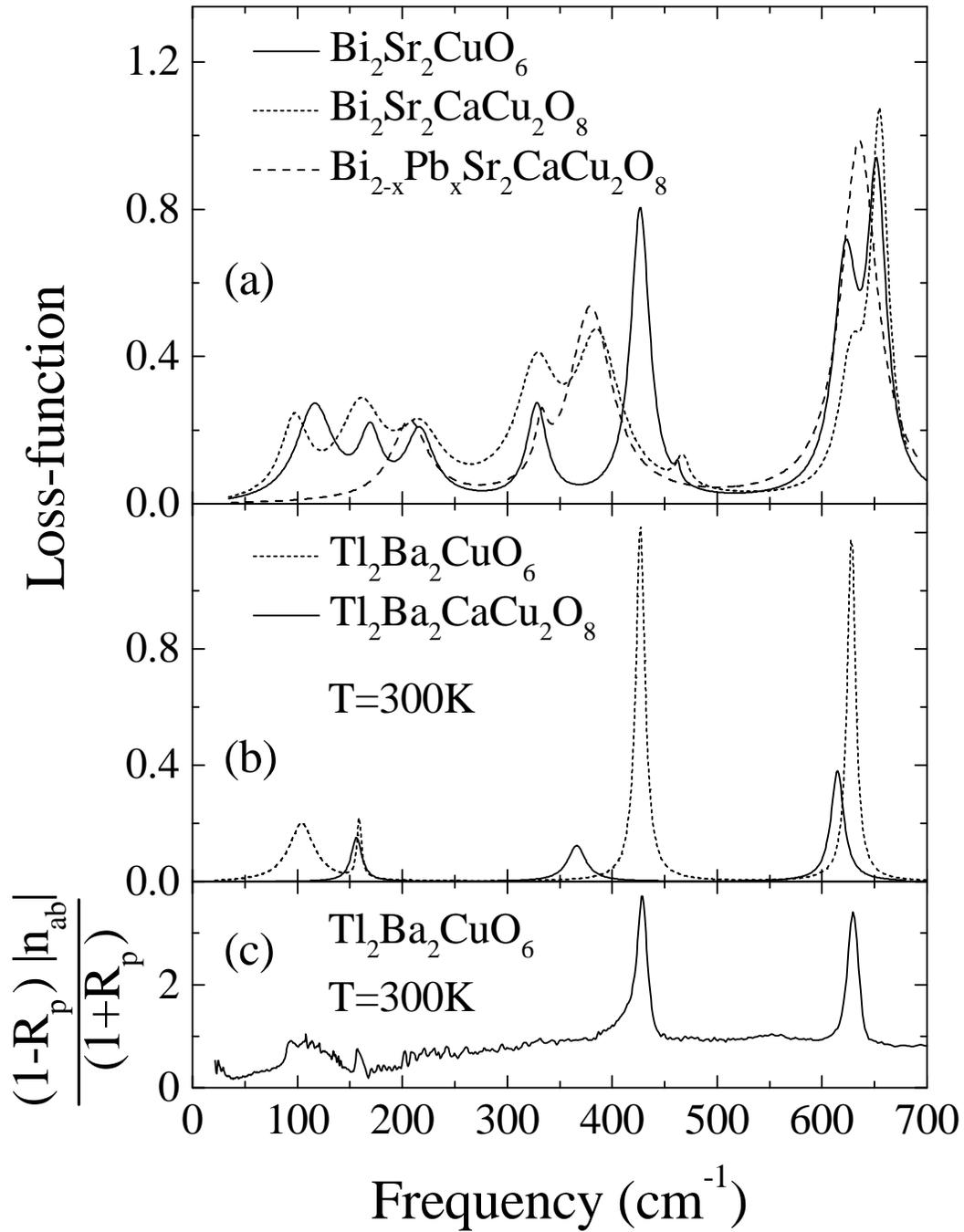}}
\vspace{20mm}
\caption{The $c$-axis loss-function (a) for the Bi and
(b) for Tl compounds, obtained from the fit of 
the reflectivity to the Drude-Lorentz model.
Shown in (c) is the experimentally measured
$(1-R_p)|n_{ab}|/(1+R_p)$ function of Tl2201, which peaks
at the longitudinal frequencies.}
\label{Lossfunction}
\end{figure}

The room temperature spectra of Bi2201, Bi2212, Bi2212+Pb, Tl2201, and Tl2212 
are shown in Fig.~\ref{Reflectivity}. 
The solid lines represent the fit, and the symbols are the experimental data.
These results were reproduced on a number of Bi and Tl samples. 
The data were collected at temperatures between 8 K and 300 K. 
The main temperature variations in $R_p$ occur due to the changes in 
the in-plane response. 
Bi compounds show more optical phonons due to the lower symmetry. 
For all five compounds, the overall reflectivity scales with the in-plane 
response. 
In general, the double layer compounds have higher conductivity in the plane 
than the corresponding single layer compounds 
and, consequently, they have higher reflectivity. 
The optical conductivity of Tl cuprates
usually is at least twice as high as that of Bi.
The reflectivity $R_p$ was fitted with Eq.~\ref{eq:fresnel} and 
Eq.~\ref{eq:difun}.
The fitting parameters: TO frequencies, oscillator strengths $S_j$, and 
the damping parameters $\gamma_j$ of the phonons, as well as LO frequencies 
(obtained as the positions of maxima in the loss-function) are listed 
in Table \ref{tlfit} and Table \ref{bifit}. 
The temperature dependence of the phonon parameters can be seen from 
the comparison to the 8 K data, shown in brackets. 
The $c$-axis conductivities and loss-functions calculated from 
the fit are shown in Fig.~\ref{Conductivity} and Fig.~\ref{Lossfunction}, 
respectively.
It should be noted that the LO frequencies obtained in our experiment 
are not sensitive to the error in the absolute reflectivity and to 
the assumed in-plane conductivity, which is also fitted, 
while the TO frequencies depend on them both. 
However, this dependence is rather weak. 
In Table \ref{bifit} we showed our values of TO and LO frequencies 
in Bi2212, together with the values obtained by 
Tajima {\em et al.}\cite{Tajima} from the normal incidence reflectivity
done on the $ac$-plane at 6 K. In the latter case, the TO frequencies 
can be evaluated directly from the reflectivity spectrum. 
The LO frequencies are calculated via Kramers-Kronig transformation. 
From this comparison, one can see that the discrepancy between the LO frequency 
values obtained with the two different methods, on different samples, is only 
few wavenumbers except for the highest frequency mode, and that 
the TO values differ less than 10 cm$^{-1}$.
This means that although we have a rather complicated situation, with the mixed 
$ab$-plane and $c$-axis response, we can get quite accurate values of 
the phonon parameters: transversal and longitudinal frequencies, 
and consequently, oscillator strength.

                        %%%%%%%%%%%%%%%%%%%%%%%%%
                        %       DISCUSSION              %
                        %%%%%%%%%%%%%%%%%%%%%%%%%

\section{Discussion}

\begin{figure}[t]
\epsfxsize=14cm
\centerline{\epsffile{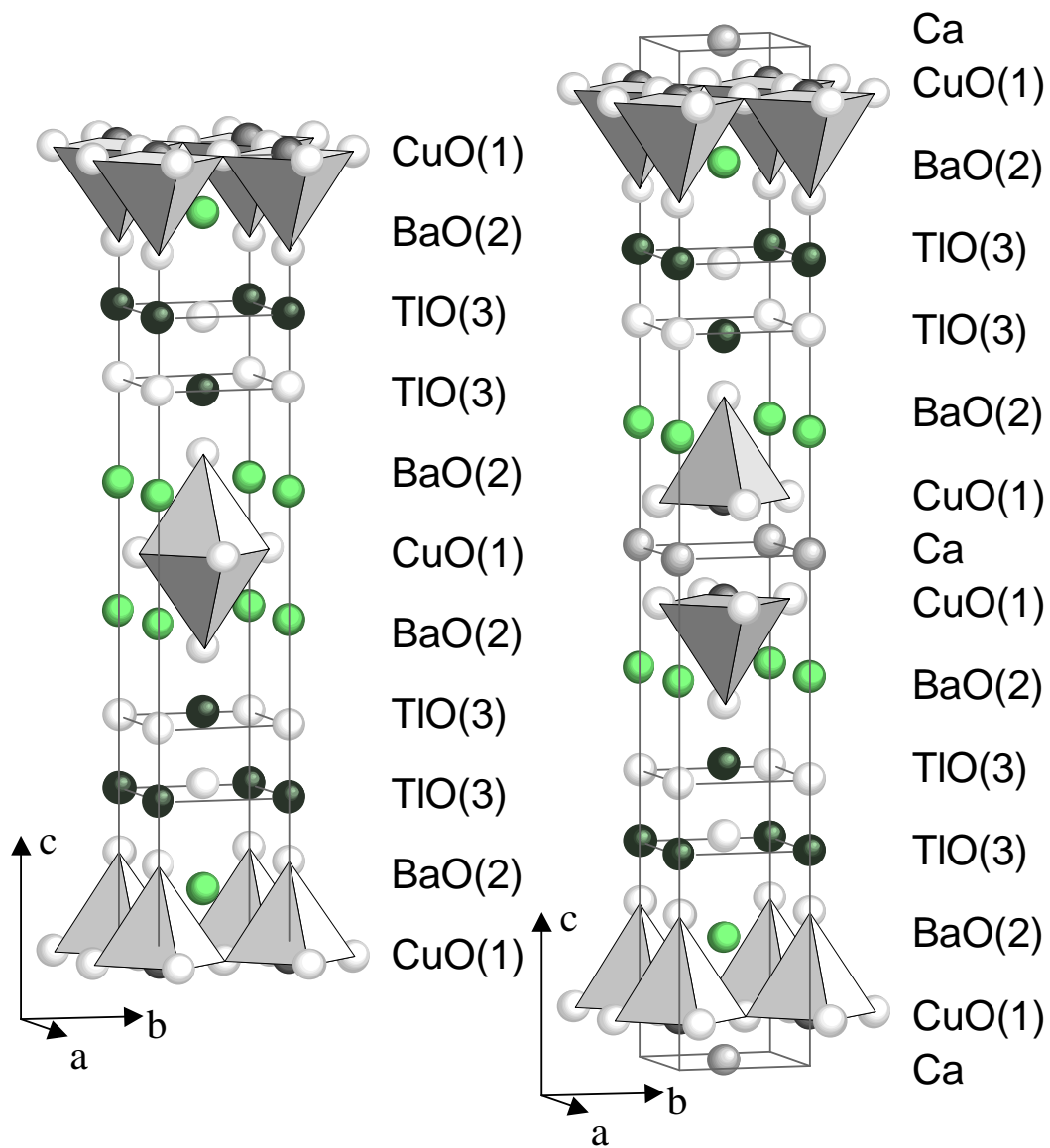}}
\vspace{20mm}
\caption{Crystal structure of Tl$_{2}$Ba$_{2}$CuO$_{6}$ (left) 
and Tl$_{2}$Ba$_{2}$CaCu$_2$O$_{8}$ (right).}
\label{Structure}
\end{figure}

A body-centered-tetragonal structure I4/{\em mmm} ($D^{17}_{4h}$) has 
been most frequently used in the interpretation of the infrared and 
Raman spectra of both bismuth and thallium cuprates. 
The unit cell for Tl$_2$Ba$_2$CuO$_6$ and Tl$_2$Ba$_2$CaCu$_2$O$_8$ is shown 
in Fig.~\ref{Structure}. The structure of Bi$_2$Sr$_2$CuO$_6$ and 
Bi$_2$Sr$_2$CaCu$_2$O$_8$ resembles the structure of the corresponding 
thallium compounds, where Bi and Sr substitute Tl and Ba, respectively.
However, the crystal lattice of superconducting Bi cuprates has orthorhombic 
distortions, which lead to the doubled $\sqrt{2}\times\sqrt{2}$ unit cell in 
the $ab$-plane. 
Moreover, there is a monoclinic incommensurate modulation of the structure with 
a period 4.5-5 {\bf b}. 
The factor group analysis for the space group I4/{\em mmm} predicts 
5 $A_{2u}$ out-of-plane and 6 $E_{u}$ in-plane IR active modes 
for single layer and $6A_{2u}+7E_{u}$ 
for double layer compounds, respectively.
However, seven $c$-axis modes can be resolved in case of Bi2201 and Bi2212 
and only four modes are clearly seen in case of Tl2201 and Tl2212 
(Fig.~\ref{Reflectivity}). 
Extra phonon modes are present in the Bi spectra due to the lower symmetry.
The six strongest $c$-axis modes are observed in the Bi2212 spectrum
at 97, 174, 220, 328, 383, and 628 cm$^{-1}$.
The last, 628 cm$^{-1}$, mode is split into main and satellite structures. 
The satellite at 655 cm$^{-1}$ and the two additional phonons 
at 469 and 472 cm$^{-1}$, which have very small optical strength and can be 
resolved only at low temperatures seem to result from orthorhombic and 
monoclinic distortions.
This implies that deviations from the basic I4/{\em mmm} structure have some, 
but rather small effect on the phonon spectrum. 

The lattice dynamical calculations for Tl2201, Tl2212 and Bi2212 were
performed using the shell-model\cite{Prade,Kulkarni} and the rigid-ion 
model.\cite{Jia1,Jia2,Jia3} The space group {\em I}4/{\em mmm} was used 
in both cases. 
The assignment of the modes and the frequencies from the lattice dynamical
calculation are given in Table \ref{phoncalc}. 
TO frequencies were calculated in the rigid-ion model, 
and TO-LO splitting from the shell-model. Although the shell-model is 
a more powerful technique, it turns out that the eigenfrequencies 
and the character of the vibrations calculated in the rigid-ion model 
are often closer to the experimental data. However, the agreement is
still not perfect. 
In our discussion, we will compare the the phonon frequencies with 
the calculated ones, but we will go beyond the I4/{\em mmm} group when 
we discuss the modulation induced modes.
From now on we will indicate the oxygen in the CuO planes as O(1), 
in the SrO (BaO) planes as O(2), and in the BiO (TlO) planes as O(3),
as shown in Fig.~\ref{Structure}.

\subsection{Phonon modes}

Let us consider first the high frequency modes. They mainly involve motion 
of light oxygen ions. 
For simplicity we will discuss here the LO frequencies of the double layer
compounds. The corresponding frequencies for the single layer compounds 
are given in Tables \ref{tlfit} and \ref{bifit}, and the assignment of the 
oxygen modes is identical for both systems. The oxygen vibration
modes are at 615cm$^{-1}$ (628/655cm$^{-1}$), 366cm$^{-1}$ (383cm$^{-1}$) and at
328 cm$^{-1}$ in the Bi compounds.

%\subsubsection{615cm$^{-1}$ mode}

\begin{figure}[t]
\epsfxsize=15cm
\centerline{\epsffile{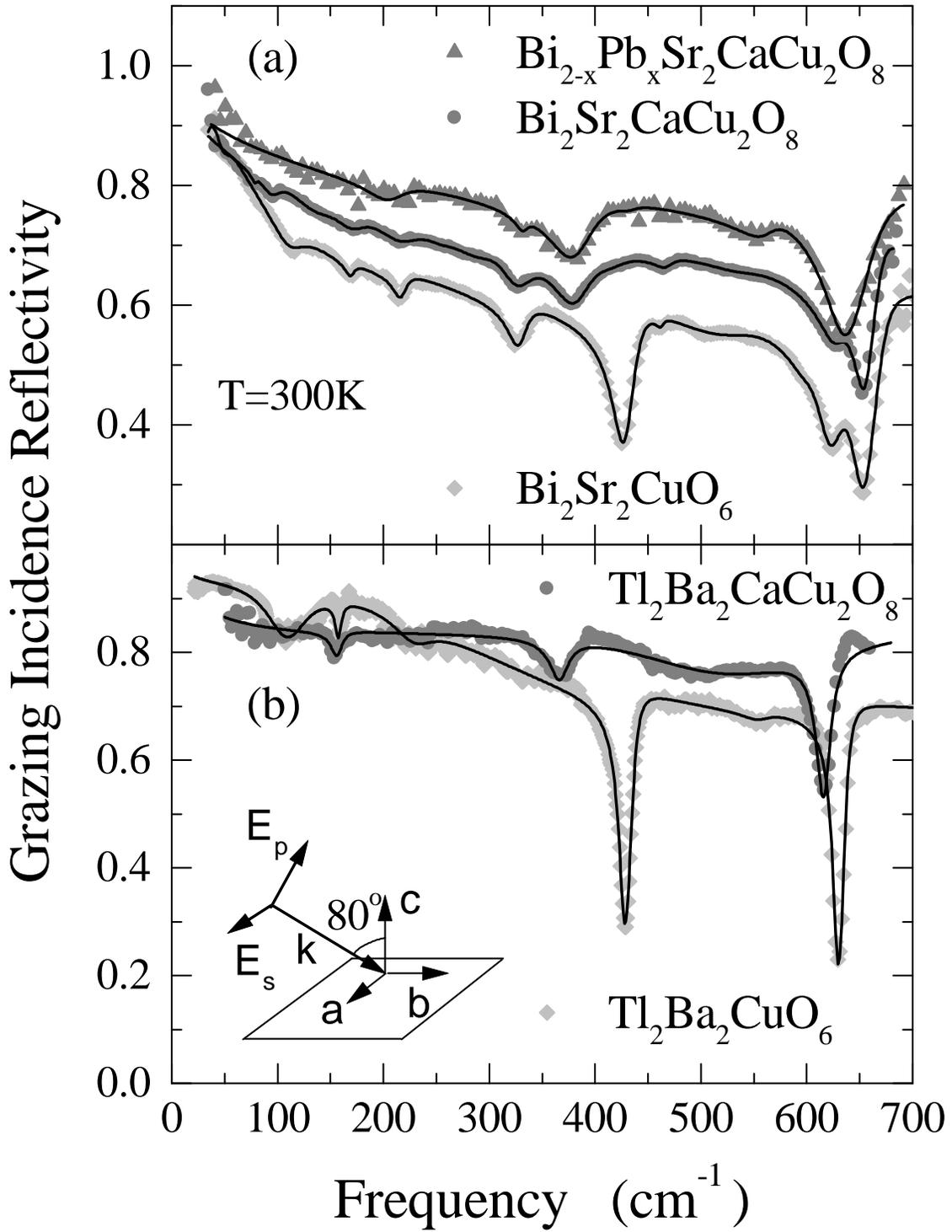}}
\vspace{20mm}
\caption{Experimental (symbols) and fitted (solid lines) grazing angle 
reflectivity $R_p$ for the p-polarized light measured 
at 80$^{\circ}$ angle of incidence at the room temperature for (a) bismuth 
and (b) thallium compounds. Sketched in the inset 
is the schematic diagram of the grazing angle technique.}
\label{Reflectivity}
\end{figure}

We assigned the highest frequency mode, i.e. at 615 cm$^{-1}$ 
(628/655 cm$^{-1}$), to the out-of-phase vibration of O(2) and O(3), 
in agreement with the lattice dynamical calculations, given in
Table \ref{phoncalc}. 
The splitting of this peak in Bi2201 and Bi2212, 
observable in Fig.~\ref{Reflectivity}a, is due to the characteristic 
incommensurate modulation\cite{Hazen} of these compounds. 
This monoclinic superstructure is very often attributed to 
the incorporation of extra oxygen atoms in the BiO planes 
every 4.5-5 periods, although O(2) vacancies or element 
substitutions are also considered to explain the superstructure. 
The modulation period 4.5 {\bf b} on our Bi2212 sample was measured by 
LEED.\cite{Karina}
Whatever the reason of the superstructure is, we use the fact that the 
modulation is not sinusoidal, but rather forms narrow stripes aligned 
along the $a$-axis, as indicated by high-resolution electron 
microscopy\cite{beskrovnyi94} and x-ray detection of higher order 
superstructure reflections. In this case, we can consider an oscillator 
model with every fifth oscillator slightly different as it could be 
the case in presence of extra oxygen. 
Because there are four undisturbed and one disturbed cells in each supercell,
the ratio of optical strength of these two modes should be around 4, 
which is close to the ratio of 4.6 observed in Bi2212 (Table \ref{bifit}).
In Bi2201, the interaction within BiO layers is 
relatively stronger, due to the weaker perovskite-like CuO$_2$ block. 
The higher interaction strength causes 
the redistribution of the spectral weight in favor of the stronger mode. 
The disappearance of the double structure in Pb doped Bi2212
(see Fig.~\ref{Reflectivity}) supports this mode assignment. 
The substitution of 20\% Bi$^{3+}$ by Pb$^{2+}$ 
relaxes the strong monoclinic superstructure distortion and induces a weaker 
orthorhombic modulation with a period at least twice as long.\cite{Zhiqiang} 
No other mode shows such noticeable splitting. Taking into account that 
the most significant displacement of ions with respect to the tetragonal 
I4/{\em mmm} structure occurs in the BiO planes and, thus, the modulation should 
mostly affect interatomic bonding in these layers, we gave the earlier mentioned 
assignment to this mode.

We ascribe the weak shallow structures between 500 and 600 cm$^{-1}$ to 
in-plane phonons. 
The minimum at 508 cm$^{-1}$ is also present in the $ab$-plane 
reflectivity of Bi2201, and it was reported as an in-plane 
phonon mode.\cite{Artem1}
A similar observation seems to hold for the 560 cm$^{-1}$ mode in 
Tl2201.\cite{Puchkov} Small sharp minima around 470 cm$^{-1}$ in Bi compounds 
are most probably zone-boundary modes, which become IR active due to the lattice 
distortion.

%\subsubsection{366 cm$^{-1}$ mode}

The next oxygen mode at 366 cm$^{-1}$ in the Tl compound 
(383 cm$^{-1}$ in Bi) is one of the strongest modes. 
It is the most noticeable feature of the spectra in 
Fig.~\ref{Reflectivity} that this mode shifts up in frequency 
by 61 cm$^{-1}$ in Tl and by 44 cm$^{-1}$ in Bi cuprates, 
going from the double to the single layer compound. 
This shift indicates that this mode involves the motion of oxygen in CuO$_2$. 
In the shell model calculations\cite{Prade,Kulkarni} (Table \ref{phoncalc}), 
the next predicted LO mode at 462 cm$^{-1}$ in Tl2212, and at 497 cm$^{-1}$
in Bi2212, is too weak and too high in frequency 
to be assigned to this vibration. The calculated mode eigenvector involves
the motion of O(1) against O(2) and Ca. However, the compound without Ca
should show a lower resonant frequency, as reported in Table \ref{phoncalc}. 
The opposite behavior of this mode indicates that maybe the influence of Ca 
on this mode is not so strong. If we consider the next available 
candidates for the assignment in Table \ref{phoncalc}, 
i.e. 294 cm$^{-1}$ in Tl2201 and 355 cm$^{-1}$ in Tl2212 modes, 
we again obtain the opposite frequency change from single layer
2201 to double layer 2212 compound. Thus, no mode predicted in 
lattice dynamical calculations could be ascribed to this phonon.
Since the observed oscillator strength of this phonon is 
very high we think that this mode involves mostly the in-phase 
motion of O(1) and O(2) against Cu. 

%\subsubsection{328 cm$^{-1}$ mode}

The mode at 328 cm$^{-1}$ is clearly seen only in the Bi compounds and it is 
present in both Bi2201 and Bi2212 at roughly the same frequency. In Tl2201, this 
mode also occurs at approximately the same frequency, but the optical strength 
is much weaker, and the line can be resolved only at a very low temperature. 
The 328 cm$^{-1}$ mode in Tl2201 shows narrowing with temperature,
which is characteristic for $c$-axis phonon lines. 
Since the frequency of the 328 cm$^{-1}$ mode in the Bi compounds does not vary 
with the number of CuO$_2$ layers, we conclude that it has O(2) 
and O(3) character. 
This phonon line is much broader than those at 427 cm$^{-1}$ in 
Bi2201, and it resembles the modulation induced splitting, which 
also supports the given assignment.

%\subsubsection{220 cm$^{-1}$ mode}

A weak structure is present in the Tl2201 
spectrum around 230 cm$^{-1}$, and it could correspond to the 220 cm$^{-1}$ 
mode in Bi compounds. 
The shape of this mode is temperature independent, which is the
characteristic behavior of in-plane phonons. We therefor attribute
this mode to an in-plane optical phonon.

The presence of the 328 cm$^{-1}$ mode with $c$-axis polarization 
and the absence for this polarization of the 220 cm$^{-1}$ mode is in 
agreement with rigid-ion calculations\cite{Jia2,Jia3} (see the 223 cm$^{-1}$ mode of 
Tl2212 in Table \ref{phoncalc} and the absence of its counterpart 
in Tl2201).
However, these experimental results contradict the shell-model 
calculations for Tl2201 (Table \ref{phoncalc}),\cite{Kulkarni} where the 355 cm$^{-1}$ 
mode is lacking. 

%\subsection{Low-frequency modes}

The low frequency modes are difficult to assign, since the heavy 
atoms involved in this kind of motion have comparable masses. 
The only mode that shifts with the change from the single to 
the double layer is the 114 cm$^{-1}$ mode in Bi2201 
(98 cm$^{-1}$ in Bi2212). 
The tendency to shift to a lower frequency is similar to 
those of the 383 cm$^{-1}$ mode, and indicates that 
the motion of Cu atoms is involved. 
The other two phonons should include the Cu-O bending 
vibrations to minimize interaction with Ca, and consequently, 
motion of Bi, Sr, Tl, and Ba atoms to preserve the center of mass. 

In all the compounds there exists a rather large discrepancy between lattice
dynamical calculations and experiment, as shown in Table \ref{phoncalc}:
The calculations predict the additional mode of O(2)-O(3) character between 
400 and 500 cm$^{-1}$ with a rather large oscillator strength. 
Experimentally, this mode seems to be absent. 
Second, the mode with the largest TO-LO splitting 
is predicted to have lower frequency in Tl2201 (294-341 cm$^{-1}$ mode)
than in Tl2212 (355-437 cm$^{-1}$ mode). The contradiction with 
the experiment can be readily observed in Fig.~\ref{Reflectivity}.
A possible reason for this inconsistency is probably the fact that 
the early lattice dynamical calculations have been adjusted to 
data obtained on polycrystalline  materials,\cite{Popovic,Foster,Zetterer,Kamaras,Mihajlovic} on which  
the in-plane and out-of-plane optical phonons can not be separated. 
In fact, in Tl2223 and Bi2212 single crystals the in-plane phonons have been 
observed at 415 cm$^{-1}$ and 420 cm$^{-1}$, respectively.\cite{shimada} 

\subsection{Resonances in intrinsic Josephson junctions}

In the superconducting state the high-T$_{\text{c}}$ compounds are considered 
to form a stack of superconducting CuO$_2$ planes separated by 
the non-superconducting heavy element rock-salt-type 
layers, where each pair of CuO$_2$ planes forms a Josephson junction. 
The interplane current showing typical Josephson behavior was observed in
$I-V$ characteristics of thin mesoscopic structures of Bi2212 and 
Tl2223.\cite{Kleiner92} 
Subsequent measurements revealed a number of small intra-gap 
resonances on the resistive branch of the $I-V$ 
curves.\cite{Schlenga2,Yurgens} The following investigations 
ruled out many known effects related to the superconducting energy gap or 
geometrical resonances. The $I-V$ characteristic of 
a stack of junctions was shown to be a mere superposition of 
$I-V$ curves of individual junctions,\cite{Schlenga} which implies 
that all junctions are independent and that the intra-gap structures occur
at characteristic bias voltages in every Josephson junction.
Recently, Helm {\it et al}.\cite{Helm} proposed that
oscillations of the electric field within a junction, excited by
the bias voltage due to the $ac$ Josephson effect,
couple to the IR active lattice 
vibrations and resonate at the longitudinal modes of the $c$-axis phonons. 
The comparison to the optical data was not done, because
the resonances lay well within the gap and the corresponding frequencies 
are difficult to access with presently available samples. 
Moreover, in the usual normal incidence technique, 
the longitudinal frequencies are obtained by
Kramers-Kronig analysis and can be biased by noise and an error 
in absolute reflectivity at low frequencies. 

\begin{figure}[t]
\epsfxsize=15cm
\centerline{\epsffile{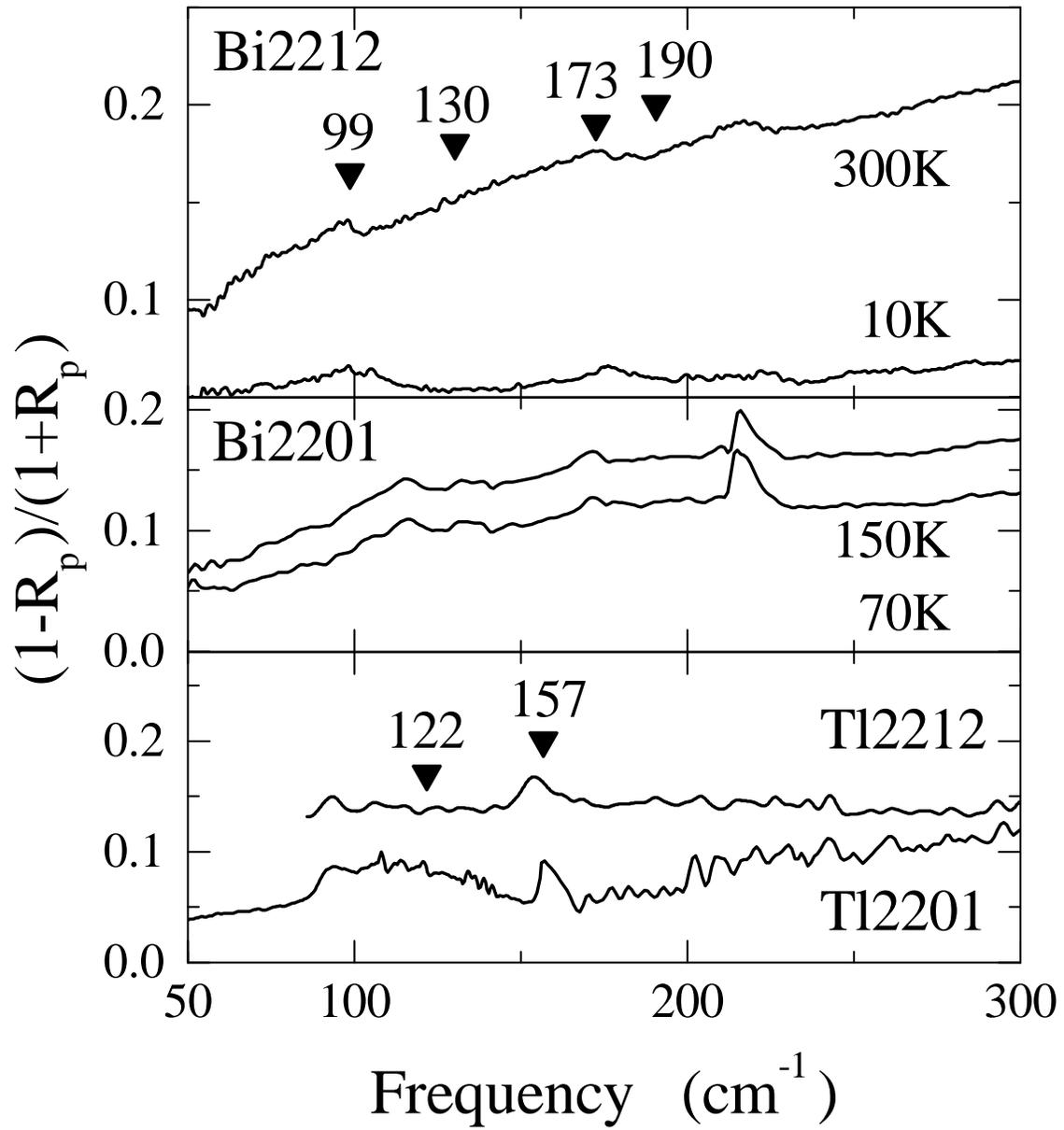}}
\vspace{20mm}
\caption{Function $(1-R_p)/(1+R_p)$ for Bi2212 (a), Bi2201 (b),
Tl2212 and Tl2201 (c) for different temperatures in comparisson
with the positions of the resonances observed by the intrinsic
Josephson junction spectroscopy: (a) for Bi2212 (marked by 
triangles and the corresponding value) and (b) for Tl2223.}
\label{IJE}
\end{figure}

In Fig.~\ref{IJE} the measured function $(1-R_p)/(1+R_p)$, where maxima 
correspond to the $c$-axis longitudinal frequencies is shown for Bi2212 
together with the positions of the four pronounced resonances observed 
in Ref.~\onlinecite{Schlenga}. In the superconducting state the phonon 
structure does not reveal substantial changes as it can be seen 
from the 10 K curve. In the lower panel of the Fig.~\ref{IJE}, 
the optical data for Tl2201 and Tl2212 are compared to 
the tunnel data for Tl2223. 
Two phonons in the Bi2212 curve are in good agreement with the $I-V$ 
resonances at 99 and 173 cm$^{-1}$, and at least one coincides in the 
case of Tl2223. 
The latter phonon is observed in Tl2201 and Tl2212, but its position 
is almost unchanged in the systems with one and two CuO$_2$ layers per
unit cell. We can expect that the third layer in Tl2223 should not
substantially shift its frequency.

\begin{figure}[t]
\epsfxsize=15cm
\centerline{\epsffile{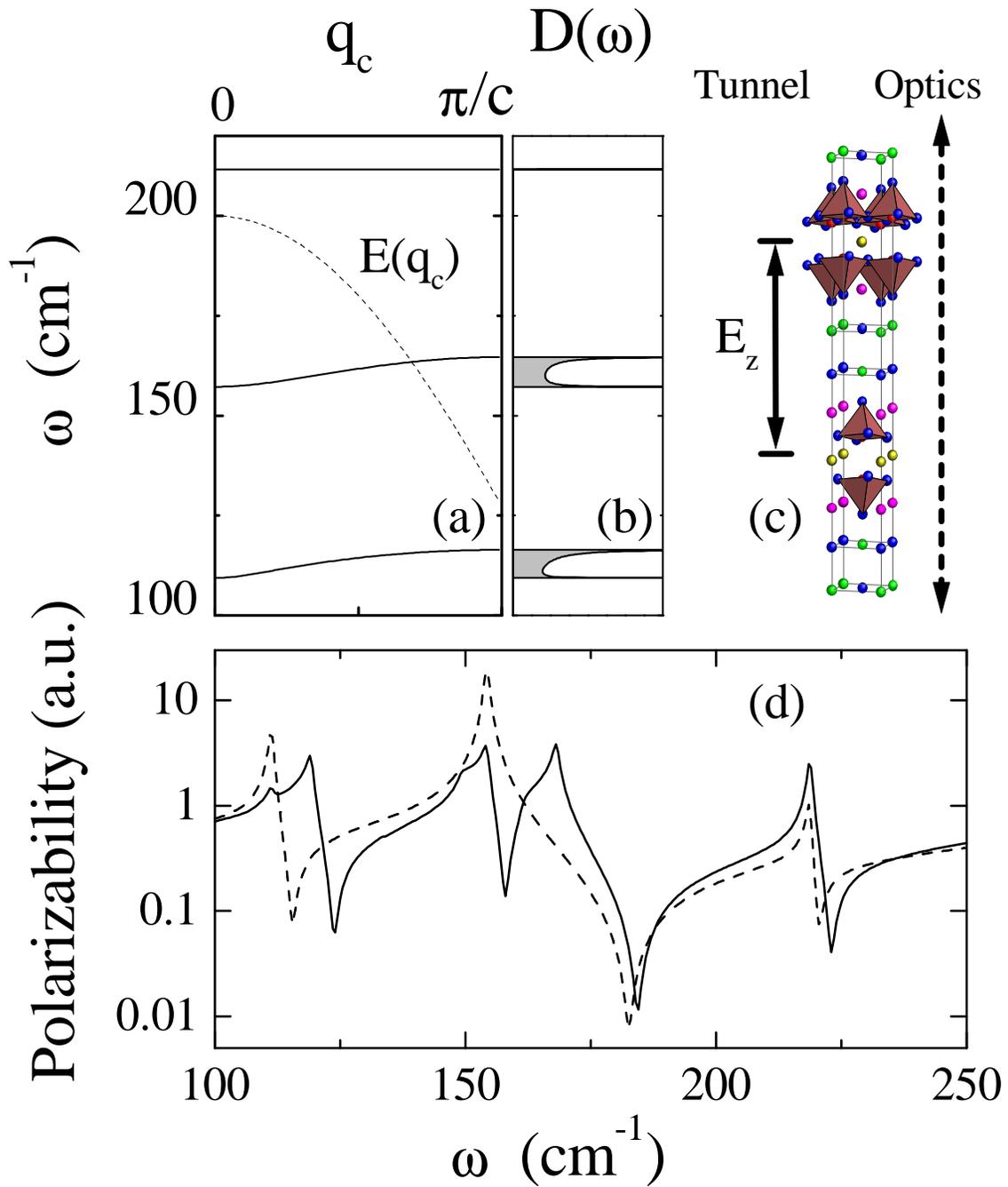}}
\vspace{20mm}
\caption{(a) Simulated phonon dispersion of three branches in 
the low frequency range for Bi2212 and $q_c$-dispersion of 
the electric field in tunneling measurements. (b) The phonon 
density of states corresponding to the dispersion curves in (a). 
(c) Tunnel vs optics electric field
distribution within the crystal. In the tunnelling measurements, the electric
field is localized within one junction. (d) Polarizability for the uniform
(dashed line) and non-uniform (solid) electric field distribution, simulated 
on the stack of the $ab$-planes, whith each plane representing
an atom in the unit cell.}
\label{Dispersion}
\end{figure}

This agreement let us conclude that, as stated in the model of 
Helm {\it et al}.,\cite{Helm} the optical phonons play 
the main role in the structure observed in the $I-V$ characteristics. 
However, more resonances are observed in the tunnelling experiment. 
The reason for this discrepancy between theory and experiment is that
Helm {\it et al}.\cite{Helm} consider the barrier filled
with the macroscopic quantity of the material with properties similar 
to the bulk.
In the following we will show that the additional resonances
can correspond to zone-boundary modes, which are excited due to 
the non-uniform field distribution in the tunnelling measurements. 
Indeed, the electric field $E$ in the interlayer Josephson 
junction oscillates only between two adjacent CuO$_2$ planes and 
vanishes outside the junction.
Thus, in the wavevector space, $E(q_c)$ spans 
the whole Brillouin zone following the Fraunhofer diffraction pattern. 
In optical measurements, the electromagnetic wave decays in the sample 
over the penetration depth length scale, which is much larger than 
the interplane distance. The corresponding distribution in $q_c$ space is 
represented by the $\delta$-function near $q_c=0$. 
The electric field in the Josephson junction excites phonons for all 
$q_c$, and the induced polarization has peaks at the maxima of 
the phonon density of states. To illustrate this qualitatively we calculated 
the phonon dispersions for the three lowest frequency modes using a reduced 
rigid-ion model. The curves are plotted in Fig.~\ref{Dispersion}a, 
together with the electric field distribution $E(q_c)$. 
For this calculation we 
represented each atom in the unit cell by a plane perpendicular to 
the $c$-axis. 
The force constants, which include the nearest and next-nearest
interactions in the unit cell are adopted from the Ref.~\onlinecite{Jia1}. 
The phonon density of states $D(\omega)$ is shown in 
Fig.~\ref{Dispersion}b. 
It can be seen that the polarization should peak at the zone-center and 
zone-boundary phonon frequencies.
Fig.~\ref{Dispersion}d shows the frequency dependence of the polarization for 
the uniform field distribution (dashed curve) and for the non-uniform 
box-car distribution (solid curve). In the latter case the electric field 
amplitude is finite (and constant) only within a single unit cell as 
shown in Fig.~\ref{Dispersion}c.
To simulate this non-uniform field distribution, we extended our model 
to 1000 unit cells within the chain, with a finite dissipation 
introduced to avoid problems due to the circular boundary conditions. 
Fig.~\ref{Dispersion}d shows that the polarization indeed has maxima 
both at $\omega(q_c=0)$ and $\omega(\pi/c)$. 

The zone-boundary phonons modes can become optically active if lattice 
distortions occur in the crystal structure. In Fig.~\ref{IJE} we plotted 
the data for Bi2201. 
The orthorhombic and monoclinic distortions of I4/{\em mmm} 
structure are stronger in Bi2201 than in Bi2212, and this can explain 
why the 130 cm$^{-1}$ mode is observed in the single layer compound.
It should be noted that the Raman active modes could also be excited in the 
intrinsic Josephson junction measurements due to the strong ($\sim10^5$ V/cm)
polarization of the media by the DC bias voltage.

\section{Conclusion}

We performed grazing incidence reflectivity measurements on 
bismuth and thallium superconducting cuprates. 
The technique was shown to permit accurate determination of 
the $c$-axis longitudinal frequencies. 
We fitted spectra with the Drude-Lorentz model and showed that 
the values of the phonon parameters agree with 
those obtained in normal incidence measurements. 
Therefore, we argue that the grazing incidence technique is 
an excellent tool for probing the $c$-axis properties in 
high-T$_{\text{c}}$'s on samples
with small size in the $c$-direction, which is often a problem 
in normal incidence measurements.

We compared the $c$-axis TO and LO phonon frequencies in five
structurally similar compounds, Tl$_{2}$Ba$_{2}$CuO$_{6}$, 
Tl$_{2}$Ba$_{2}$CaCu$_{2}$O$_{8}$, 
Bi$_{2}$Sr$_{2}$CuO$_{6}$,
Bi$_{2}$Sr$_{2}$CaCu$_{2}$O$_{8}$, and
Bi$_{2-\text{x}}$Pb$_{\text{x}}$Sr$_{2}$CaCu$_{2}$O$_{8}$, 
with lattice dynamical calculations and proposed a new assignment for 
the main oxygen modes. 
The major difference is the assignment of the 366 cm$^{-1}$ mode in 
2212 compounds, which we consider to be the in-phase
motion of O(1) and O(2) against Cu. We observed a substantial discrepancy between
the available lattice dynamical calculations and our experimental data:
The calculations predict a phonon mode between 400 and 500 cm$^{-1}$, which
we do not observe. Shell model calculations indicate that 
the oxygen mode with the largest TO-LO splitting in the double 
layer 2212 compounds should be higher in frequency than those in 
the single layer 2201 compounds, which is again inconsistent 
with our data.

The comparison of the LO frequencies obtained by our method with those inferred 
from intrinsic Josephson effect measurements shows that the latter 
technique - due to the non-uniform electric field distribution - is sensitive 
both to zone-center and zone-boundary phonons.

\acknowledgments

We thank Superconductor Technologies Inc. for providing 
the Tl2212 samples and B. A. Willemsen for his assistance.
We gratefully acknowledge useful discussions with W. N. Hardy, P. M\"uller, 
R. Kleiner, and Ch. Helm.
We would like to thank H. Bron for performing chemical microanalysis of samples.
This investigation was supported by the Netherlands Foundation for 
Fundamental Research on Matter (FOM) with financial aid from 
the Nederlandse Organisatie voor Wetenschappelijk Onderzoek (NWO). 
A.A.T, G.A.K., J.I.G, and N.N.S. acknowledge support from 
the Russian Superconductivity Program under 
project \# 96-120 and RFBR under project \# 97-02-17593.
The work performed at SUNY-Buffalo was supported by US
Department of Energy grant No. DE-FG02-98ER45719

%%%%%%%%%%%%%%%%%%%%%%%%%%%
%    The references       %
%%%%%%%%%%%%%%%%%%%%%%%%%%%

\newpage

        %%%%%%%%%%%%%%%%%%%%%%%%%
        %       TABLES          %
        %%%%%%%%%%%%%%%%%%%%%%%%%

%%%%table 2 Tl2201, Tl2212 data
%%%%%%
\begin{table}
\caption{$c$-axis phonon parameters of the Tl2201 and Tl2212 at room 
temperature.
The values between the brackets were obtained at 8 K, $\varepsilon_\infty=4$.}
\begin{tabular}{cccc}
\multicolumn{4}{c}{Tl$_{2}$Ba$_{2}$CuO$_{6}$} \\ 
\tableline
    $\tilde{\nu}_{T_{j}}$  $[cm^{-1}]$ & 
$\tilde{\nu}_{L_{j}}$  $[cm^{-1}]$ & $\gamma_j$ $[cm^{-1}]$ & $S_j$ \\
\tableline 
77.9 (77.9) & 111.0 (111.3) & 26.5 (26.5) & 4.5 (4.5)\\
156.5 (156.0) & 158.0 (157.8)  & 8.6 (6.1) & 0.12 (0.11) \\ 
385.5 (385.7) & 427.1 (426.7) & 8.3 (6.5) & 0.94 (0.95) \\ 
603.2 (603.3) & 628.3 (628.5) & 8.7 (8.6) & 0.28 (0.28) \\ \hline
\multicolumn{4}{c}{Tl$_{2}$Ba$_{2}$CaCu$_{2}$O$_{8}$} \\ \hline
$\tilde{\nu}_{T_{j}}$  $[cm^{-1}]$ & 
$\tilde{\nu}_{L_{j}}$  $[cm^{-1}]$ & $\gamma_j$ $[cm^{-1}]$ & $S_j$ \\ \hline
152.0 (152.0) & 156.4 (157.0) & 12.2 (9.0) & 0.13 (0.13) \\ 
339.0 (339.0) & 366.3 (367.0) & 20.8 (18.0) & 1.02 (1.10)\\ 
598.1 (602.0) & 614.7 (617.0) & 17.1(12.0) & 0.22 (0.23) \\ 
\end{tabular}
\label{tlfit}
\end{table}

%%%%%%%%%%%%%%%%%%%%%%%%
%TABLE
%%%%%%%%%%%%%%%%%%%%%%%%
\begin{table}
\caption{$c$-axis phonon parameters of the Bi2201 and Bi2212
at 300K (8K), $\varepsilon_\infty=4$.}
\begin{tabular}{cccccc}
\multicolumn{6}{c}{Bi$_{2}$Sr$_{2}$CuO$_{6}$} \\ 
 \tableline
\multicolumn{2}{c}{$\tilde{\nu}_{T_{j}}$  $[cm^{-1}]$} & 
\multicolumn{2}{c}{$\tilde{\nu}_{L_{j}}$  $[cm^{-1}]$} & 
$\gamma_j$ $[cm^{-1}]$ & $S_j$ \\ 
 \tableline
\multicolumn{2}{c}{83.0 (87.0)} & \multicolumn{2}{c}{118.3 (115.6)} & 15.0 
(12.0) & 0.35 (0.36) \\
\multicolumn{2}{c}{144.5 (148.7)} & \multicolumn{2}{c}{169.7 (175.0)} & 11.0 
(11.0) & 0.70 (0.68) \\ 
\multicolumn{2}{c}{196.4 (198.3)} & \multicolumn{2}{c}{217.2 (218.0)} & 16.0 
(12.1) & 0.68 (0.60) \\ 
\multicolumn{2}{c}{292.0 (307.4)} & \multicolumn{2}{c}{328.4 (329.8)}& 25.0 
(21.7) & 1.20 (1.02) \\ 
\multicolumn{2}{c}{381.0 (390.5)} & \multicolumn{2}{c}{426.8 (427.0)}& 12.5 
(10.8) & 0.65 (0.62) \\ 
\multicolumn{2}{c}{461.6 (461.8)} & \multicolumn{2}{c}{462.6 (462.8)}& 10.0 (8.0) 
 & 0.002 (0.002) \\ 
\multicolumn{2}{c}{583.8 (595.7)} & \multicolumn{2}{c}{622.5 (626.1)}& 24.5 
(22.0) & 0.46 (0.41) \\ 
\multicolumn{2}{c}{638.9 (643.8)} & \multicolumn{2}{c}{652.2 (654.5)} & 16.4 
(11.5) & 0.04 (0.04)\\ \hline 
\multicolumn{6}{c}{Bi$_{2}$Sr$_{2}$CaCu$_{2}$O$_{8}$} \\
 \tableline
$\tilde{\nu}_{T_{j}}$ & $\tilde{\nu}_{L_{j}}$ &
$\gamma_j$ & $S_j$ & 
$\tilde{\nu}_{T_{j}}$\tablenotemark[1]
& $\tilde{\nu}_{L_{j}}$\tablenotemark[2] \\ 
 \tableline
81.0 & 96.7 & 15.0 & 0.80 &  & \\ 
(77.9) & (94.8) & (12.0) & (0.76) & 95 & 97 \\ 
157.2 & 173.5 & 34.4 & 0.59 &  & \\ 
(160.0) & (177.4) & (30.1) & (0.32) & 173 & 175 \\ 
198.3 & 220.0 & 35.3 & 0.79 &  & \\ 
(200.0) & (222.0) & (35.0) & (0.70) & 210 & 218 \\ 
301.2 & 328.4 & 33.0 & 0.88 &  & \\ 
(325.0) & (330.6) & (16.2) & (0.98) & 312 & 327 \\ 
356.9 & 383.0 & 49.7 & 0.21 &  &  \\ 
(365.0) & (372.9) & (28.9) & (1.02) & 359 & 380 \\ 
461.7 & 468.7 & 7 & 0.001 &  & \\
(465.0) & (467) & (2.1) & (0.001) & 468 & 470 \\
470.2 & 472.3 & 10 & 0.003 &   &   \\
(478.0) & (480) & (5.0) & (0.003) &   &   \\
610.0 & 627.7 & 28.7 & 0.23 &  &  \\
(613.5) & (631.1) & (27.0) & (0.22) & 580 & 630 \\
638.5 & 655.1 & 13.0 & 0.05 &  & \\ 
(646.9) & (655.9) & (12.9) & (0.05) & 622 & 675 \\ 
 \tableline
\multicolumn{6}{c}{Bi$_{2-x}$Pb$_x$Sr$_{2}$CaCu$_{2}$O$_{8}$} \\ 
 \tableline
\multicolumn{2}{c}{$\tilde{\nu}_{T_{j}}$ $[cm^{-1}]$} & 
\multicolumn{2}{c}{$\tilde{\nu}_{L_{j}}$ $[cm^{-1}]$} & 
$\gamma_j$ $[cm^{-1}]$ & $S_j$ \\ 
 \tableline
\multicolumn{2}{c}{192.8}  & \multicolumn{2}{c}{208.0}  & 43.8 & 1.22 \\ 
\multicolumn{2}{c}{305.3}  & \multicolumn{2}{c}{332.5}  & 15.8 & 1.37 \\ 
\multicolumn{2}{c}{350.0}  & \multicolumn{2}{c}{380.2}  & 29.8 & 0.34 \\
\multicolumn{2}{c}{593.5}  & \multicolumn{2}{c}{635.7}  & 28.1 & 0.72 \\
\end{tabular}
\tablenotetext[1]{Positions of maxima in $\varepsilon^{\prime\prime}$, 
from Tajima {\it et al}.\cite{Tajima}}
\tablenotetext[2]{Positions of maxima in the loss-function, 
from Tajima {\it et al}.\cite{Tajima}}
\label{bifit}
\end{table}

\begin{table}
\caption{Calculated values of the $c$-axis phonon parameters}
\begin{tabular}{ccccc}
This & 
\multicolumn{2}{c}{Prade {\em et al.},\cite{Prade}} & 
\multicolumn{2}{c}{Jia {\em et al.}\cite{Jia1,Jia2,Jia3}} \\ 
work& 
\multicolumn{2}{c}{Kulkarni {\em et al.}\cite{Kulkarni}} & 
\multicolumn{2}{c}{ } \\ 
 \tableline
$\nu_{T}$ & Assignment & $\tilde{\nu}_{T_{j}}$ 
($\tilde{\nu}_{L_{j}}$) & Assignment & $\tilde{\nu}_{T_{j}}$ \\ 
 \tableline
\multicolumn{5}{c}{Tl$_{2}$Ba$_{2}$CuO$_{6}$} \\ 
 \tableline
78 & 
Cu, Ba, Tl$^\prime$ & 108 (108) &
Cu, Tl$^\prime$ & 146  \\ 
157 & 
Cu, Ba$^\prime$ & 129 (143) & Cu, Tl, Ba$^\prime$ &160 \\ 
386 & O(1) & 294 (341) & O(1), Ba$^\prime$ & 376 \\ 
 & O(2), O(1)$^\prime$ & 413 (451) & O(2), O(3), Tl$^\prime$  & 537  \\ 
603 & O(3), O(2)$^\prime$ & 602 (648) & O(3), O(2)$^\prime$ & 568 \\ 
 \tableline
\multicolumn{5}{c}{Tl$_{2}$Ba$_{2}$CaCu$_{2}$O$_{8}$} \\ 
 \tableline
 & Cu, Ba, Tl$^\prime$ & 108 (108) & Cu, Ba, Tl$^\prime$ &95  \\ 
152 & Cu, Ba$^\prime$ & 113 (134) & Cu, Ba$^\prime$ &132  \\ 
 & O(1),Ca & 210 (238) & O(1), Ca, Ba$^\prime$ &223  \\ 
339 & O(2), O(3), Ca$^\prime$ &355 (437) & Ca, Ba, O(1), O(2)  &371  \\ 
 & O(2), O(1)$^\prime$, Ca & 457 (462) & O(2), O(3) &500  \\ 
598 & O(3), O(2)$^\prime$ & 591 (630) & O(2), O(3)$^\prime$ & 568 \\ 
 \tableline
\multicolumn{5}{c}{Bi$_{2}$Sr$_{2}$CaCu$_{2}$O$_{8}$} \\ 
 \tableline
81 & & & & \\
157 & Cu, Sr, Bi$^\prime$ & 137 (141) & Cu, Sr, Bi$^\prime$ &164  \\ 
198 & Cu, Sr$^\prime$ & 169 (169) &Cu, Sr$^\prime$  &225  \\ 
301 & O(1), Ca, Sr$^\prime$ & 277 (278) &O(1), Ca, Sr$^\prime$  &281  \\ 
357 & O(3), Ca$^\prime$ & 334 (477) &O(1), O(2)  & 442 \\ 
(462) & O(1), O(3)$^\prime$, Ca & 487 (497) &O(2), O(3) &467  \\ 
610 & O(3), O(2)$^\prime$ & 514 (530) &O(2), O(3)$^\prime$ & 617 \\ 
\end{tabular}
\label{phoncalc}
\end{table}

\newpage
\onecolumn

        %%%%%%%%%%%%%%%%%%%%%%%%%
        %       FIGURES         %
        %%%%%%%%%%%%%%%%%%%%%%%%%


%%%%%%%%%%%%%%%%%%%%%%%%%%%
%    The figures          %
%%%%%%%%%%%%%%%%%%%%%%%%%%%

%%%%%%%%%%%%%%%%%%%%%
% Replace xxx.eps with the filename of your postscript figure
% Replace xxxlabel with e.g. 'anything' else. You can refer to this 
% figure in the text with \ref{anything}.
%%%%%%%%%%%%%%%%%%%%
%%%%%%%%%%%%%%%%%%%%%
%\begin{figure}
%\centerline{\epsfig{figure=paper2.eps,width=7cm,clip=}}
 % \caption{Grazing incidence reflectivity of Bi2201, Bi2212 (upper panel),
  %and Tl2201, Tl2212 (lower panel) at T=300K. Inset: the geometry of the experiment.}
 % \label{paper2}
%\end{figure}
%\begin{figure}
%\centerline{\epsfig{figure=loss2.eps,width=7cm,clip=}}
%\caption{$c$-axis loss-function of Tl2201, Tl2212 (upper panel), and Bi2201, Bi2212 (lower panel).}
%\label{loss2}
%\end{figure}
%\begin{figure}
%\centerline{\epsfig{figure=sig2.eps,width=7cm,clip=}}
%\caption{$c$-axis conductivity of Tl2201, Tl2212 (upper panel), and Bi2201, Bi2212 (lower panel).}
%\label{sig2}
%\end{figure}
\end{document}